\documentclass[final,5p,number,times,twocolumn]{elsarticle}
\usepackage{graphicx}

\newcommand{\lr}[1]{ \left( #1 \right) }
\newcommand{\lrs}[1]{ \left[ #1 \right] }

\newcommand{\vev}[1]{ \langle \, #1 \, \rangle }

\newcommand{\rvac}{ \, | 0 \rangle }
\newcommand{\lvac}{ \langle 0 | \, }

\newcommand{\expa}[1]{ \exp{\left( #1 \right)} }

\journal{Physics Letters B}

\begin{document}
\sloppy
\begin{frontmatter}

\title{Numerical study of chiral symmetry breaking\\
in non-Abelian gauge theory with background magnetic field}

\author[JIPNR,ITEP]{P. V. Buividovich}
\author[LMPT,DMPA,ITEP]{M. N. Chernodub}
\author[ITEP]{E. V. Luschevskaya}
\author[ITEP]{M. I. Polikarpov}
\address[JIPNR]{JIPNR ``Sosny'', National Academy of Science, Acad. Krasin str. 99, Minsk, 220109 Belarus}
\address[ITEP]{Institute for Theoretical and Experimental Physics, B. Cheremushkinskaya 25, Moscow, 117218 Russia}
\address[LMPT]{Laboratoire de Mathematiques et Physique Theorique,
CNRS UMR 6083, F\'ed\'eration Denis Poisson, Universit\'e de Tours, Parc de Grandmont, 37200, France}
\address[DMPA]{Department  of Mathematical Physics and Astronomy, University of Gent, Krijgslaan 281, S9, B-9000 Gent, Belgium}

\date{November 9, 2009}
\begin{abstract}
We investigate the effect of a uniform background magnetic field on the chiral symmetry breaking in
$SU(2)$ Yang--Mills theory on the lattice. We observe that the chiral condensate grows linearly with
the field strength $B$ up to $\sqrt{e B} = 3 \, \mbox{GeV}$ as predicted by chiral perturbation theory
for full QCD. As the temperature increases the coefficient in front of the linear term gets smaller.
In the magnetic field near-zero eigenmodes of the Dirac operator tend to have more regular structure
with larger (compared to zero-field case) Hausdorff dimensionality. We suggest that the delocalization
of near-zero eigenmodes plays a crucial role in the enhancement of the chiral symmetry breaking.
\end{abstract}


\end{frontmatter}

\section{Introduction}

Electromagnetic interactions are most commonly used to probe various
properties of other fundamental interactions, in particular, of the
strong interactions which bind quarks into hadrons. In many
physically relevant situations, which can be encountered, for
example, in astrophysics or nuclear collisions, a reasonable
approximation is that of a constant background electric or magnetic
field. For instance, very strong magnetic fields ($B \sim 10^{16} \,
\mbox{T}$, $\sqrt{e B} \sim 1 \, \mbox{GeV}$) could have been
produced in the early Universe after the electroweak phase
transition \cite{Vachaspati:91:1}. At the present age of the
Universe, the strongest magnetic fields can probably be generated in
compact dense stars, such as magnetars ($B \sim 10^{10} \,
\mbox{T}$, $\sqrt{e B} \sim 1 \, \mbox{MeV}$) \cite{Duncan:92:1}. A
controllable way to subject hadronic matter to a strong
electromagnetic field may be provided by heavy-ion collisions in
very strong laser pulses. The latter possibility will be soon
realized in such experiments as PHELIX at GSI \cite{PHELIX} and ELI
\cite{ELI}, with estimated values of $\sqrt{e B}$ of order $0.01 \,
\mbox{MeV}$ ($B \sim 10^{7}\, \mbox{T}$ at the intensity of laser
radiation $I \sim 10^{23}\, \mbox{W}/\mbox{cm}^{2}$). Thus
experimentally available field strengths are still much smaller than
hadronic scale.

However, very strong electromagnetic fields with strength comparable
to hadronic scale ($B \sim 10^{15} \, \mbox{T}$, $\sqrt{e B} \sim
300 \, \mbox{MeV}$) can be produced in noncentral heavy-ion
collisions \cite{Rafelski:78:1,Kharzeev:08:1,Selyuzhenkov:06:1}.
In this case, the magnetic field perpendicular to the reaction plane
is due to the relative motion of the ions themselves. Such a strong
magnetic field breaks $P$ and $CP$ symmetries and leads to charge
separation, which manifests itself in a specific correlations
between emitted positively and negatively charged particles
\cite{Kharzeev:08:1,Selyuzhenkov:06:1}. Magnetic fields of the same
order should also modify significantly the properties of QCD vacuum
and can even change the order of the phase transition to quark-gluon
plasma \cite{Agasian:Fraga:08:1:2:3}. In particular, magnetic field
couples differently to particles with different chiralities enhancing the spontaneous breaking of the chiral
symmetry of QCD, (i.e., the symmetry with respect to the interchange
of massless quarks with opposite chiralities \cite{Agasian:Fraga:08:1:2:3,Gusynin:95:1,Smilga:97:1,Cohen:2007bt,Ebert:99:1,Zayakin:08:1}). This
phenomenon is now actively discussed in the literature
\cite{Agasian:Fraga:08:1:2:3,Kharzeev:08:1,Zayakin:08:1} in
connection with the heavy-ion experiments at RHIC and LHC. Another
interesting effect of strong external fields is the enhancement of
the trace anomaly~\cite{Rafelski:08:1:2}.

From a theorist's point of view, the effect of a strong background magnetic fields
on QCD vacuum is a good test for various model descriptions of strong interactions.
It turns out that magnetic field enhances the breaking of the chiral symmetry. A commonly
used order parameter for the chiral symmetry breaking is the chiral condensate
\begin{eqnarray}
\Sigma \equiv - \lvac \bar{q} q \rvac\,,
\label{eq:Sigma}
\end{eqnarray}
which vanishes if the chiral symmetry is unbroken. If the strength of magnetic field is larger than the pion mass (which is zero for massless quarks) but is still much smaller than the hadronic scale, one can use the chiral perturbation theory to calculate the field dependence of the chiral condensate \cite{Smilga:97:1}. To the lowest order in the magnetic field, the chiral condensate $\Sigma\lr{B}$ rises linearly with the field strength $B$:
\begin{eqnarray}
\label{cc_vs_B_chPT}
 \Sigma\lr{B} = \Sigma\lr{0}\, \lr{1 + \frac{e B \ln{2}}{16 \pi^{2} F_{\pi}^{2}} }\,,
\end{eqnarray}
where $F_{\pi} \approx 130 \, \mbox{MeV}$ is the pion decay
constant. The result (\ref{cc_vs_B_chPT}) is specific to QCD: for instance, in the
Nambu-Jona-Lasinio model or in AdS/QCD dual models the chiral
condensate rises quadratically with the field strength~\cite{Ebert:99:1,Zayakin:08:1,Klevansky:89:1}.

There are also other effects of the strong magnetic fields on the QCD vacuum. In a separate paper~\cite{Buividovich:2009ih}
we study the polarization of the spins of the virtual chiral quarks by the strong external field. The effect
of polarization is described by the chiral magnetic susceptibility, which is related to phenomenologically interesting
radiative transitions in hadron physics, to lepton pair photoproduction through specific chiral-odd coupling of a photon
to quarks, and to radiative heavy meson decays~\cite{ref:phenomenology}.
The chiral magnetization of the QCD vacuum includes two effects of the strong magnetic field:
the paramagnetic polarization of the quarks' spins and the enhancement of the chiral condensate.
Our paper~\cite{Buividovich:2009ih} is devoted to the spin effects, while the present article reports
our study of the chiral enhancement.

Below we describe our results of numerical studies of chiral
symmetry breaking in $SU\lr{2}$ lattice Yang-Mills theory exposed to
a uniform background magnetic field. Since non-Abelian gauge fields
interact with Abelian ones only via interaction with charged quarks,
in our quenched simulations we change only the Dirac operator
leaving the path integral weight  for non-Abelian gauge fields
intact. Note that the chiral condensate in quenched theory grows
logarithmically with lattice volume~\cite{Damgaard:2001xr} unlike the
condensate in the theory with dynamical fermions. Thus, strictly speaking,
in the thermodynamic limit the chiral condensate in the quenched theory is
infinite. We argue in Section~\ref{sec:logarithmic} that these logarithmic corrections do not affect the slope
of the linear dependence of the chiral condensate on the magnetic field~(\ref{cc_vs_B_chPT})
in the quenched limit. Thus, our result should have a smooth interpolation from
the quenched theory towards full QCD.

\section{Numerical results}

\subsection{Chiral condensate}
\label{sec:logarithmic}

In order to calculate the chiral condensate we use the Banks-Casher formula \cite{Banks:80:1},
which relates the condensate~(\ref{eq:Sigma}) with the density of near-zero eigenvalues of the Dirac
operator $\mathcal{D} = \gamma^{\mu} \, \lr{\partial_{\mu} - i A_{\mu}}$:
\begin{eqnarray}
\label{BanksCasher}
\Sigma = \lim \limits_{\lambda \rightarrow 0} \lim \limits_{V \rightarrow \infty} \, \frac{\pi \rho\lr{\lambda}}{V}
\end{eqnarray}
where $V$ is the total four-volume of Euclidean space-time.
The eigenvalues $\lambda_{n}$, the eigenmodes $\psi_{n}$ and the density of eigenvalues $\rho\lr{\lambda}$ of the Dirac operator are defined by
$$
\mathcal{D} \psi_{n} = \lambda_{n} \psi_{n}\,,\quad \rho\lr{\lambda} = \vev{ \sum \limits_{n} \delta\lr{\lambda - \lambda_{n}} }\,,
$$
where in the quenched approximation the averaging is performed over the gauge fields $A_{\mu}$ with the weight $\expa{ - S_{\mathrm{YM}}\lrs{A_{\mu}}}$ and $S_{\mathrm{YM}}\lrs{A_{\mu}}$ is Yang-Mills action. In order to implement chirally symmetric massless fermions on the lattice, we use Neuberger's overlap Dirac operator \cite{Neuberger:98:1}. Ultraviolet lattice artifacts are reduced with the help of the tadpole-improved Symanzik action for the gluon fields (see, e.g., Eq.~(1) in \cite{Luschevskaya:08:1}). Uniform magnetic field $B$ in the direction $\mu = 3$ is introduced into the Dirac operator by substituting $su\lr{2}$-valued vector potential $A_{\mu}$ with $u\lr{2}$-valued field:
\begin{eqnarray}
\label{magnetic_field}
A_{\mu}^{ij} \rightarrow A_{\mu}^{ij} + A_{\mu}^{\lr{B}} \delta^{ij}\,, \qquad
A_{\mu}^{\lr{B}} = B/2 \lr{x_{2} \delta_{\mu1} - x_{1} \delta_{\mu2}}.
\end{eqnarray}

The expression (\ref{magnetic_field}) is valid in the infinite volume. In order to combine it with periodic boundary conditions on the lattice we have introduced an additional $x$-dependent boundary twist for fermions \cite{Wiese:08:1}. In a finite hypercubic volume $L^4$ with periodic boundary conditions the total magnetic flux trough any two-dimensional face of the hypecube is quantized~\cite{Damgaard:88:1}. The uniform magnetic field takes quantized values either,
\begin{eqnarray}
q B = 2 \pi \, k/L^{2}\,, \qquad k \in \mathbb{Z}\,,
\end{eqnarray}
where
\begin{eqnarray}
q = 1/3\, e
\end{eqnarray}
is smallest (absolute value of) electric charge of the quark.

In our zero-temperature simulations we use two lattice volumes $14^{4}$ and $16^{4}$ and two lattice spacings
$a = 0.103 \, \mbox{fm}$ and $a = 0.089 \, \mbox{fm}$ in order to study finite-volume and finite-spacing effects.
We also make simulations at nonzero temperature, $T=0.82 T_c$, taking the asymmetric lattice $16^3 \times 6$.
The critical temperature in $SU(2)$ gauge theory is $T_c = 313.(3)$~MeV~\cite{ref:Vitaly}. The parameters of
our simulations and minimal nonzero values of the magnetic field are summarized in Table~\ref{tbl:parameters}.
\begin{table*}
\begin{center}
\caption{Parameters of simulations and fitting results for the chiral condensate.}
\label{tbl:parameters}
\begin{tabular}{cccccccc|ll}
\hline
\hline
\multicolumn{8}{c}{Parameters of simulations} &
\multicolumn{2}{|c}{Fitting results} \\
\hline
$T/T_c$ & $L_s$ & $L_t$ & $\beta$  & $a$, fm  & $L_s a$, fm  & $V_{4D}$, fm$^4$ & $ \sqrt{e B_\mathrm{min}}$, MeV & $\Sigma_0^{1/3}$, MeV & $\Lambda_B$, GeV \\
\hline
0     &  14    & 14   & 3.2810   & 0.103     & 1.44 & 4.3  &  593  &  307(15)  & 1.36(13)  \\
0     &  16    & 16   & 3.2810   & 0.103     & 1.65 & 7.4  &  519  &  320(5)   & 1.53(11)  \\
0     &  16    & 16   & 3.3555   & 0.089     & 1.42 & 4.1  &  601  &  331(19)  & 1.43(21)  \\
0.82  &  16    &  6   & 3.1600   & 0.128     & 2.01 & 6.6  &  418  &  291(1)   & 1.74(3) \\
\hline
\hline
\end{tabular}
\end{center}
\end{table*}

\begin{figure}[ht]
\begin{center}
  \includegraphics[width=6cm, angle=-90]{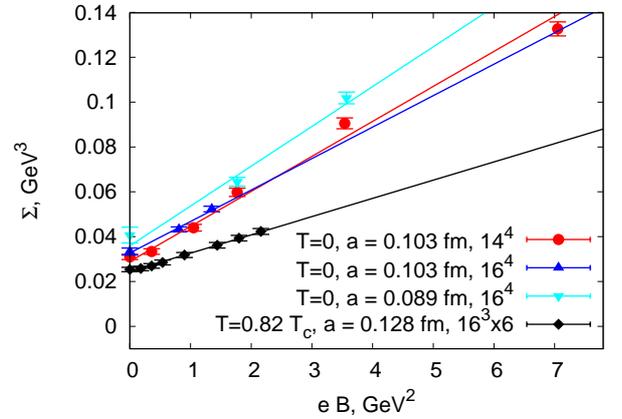}
\end{center}
  \caption{The dependence of the chiral condensate on the strength of external magnetic field at different lattice volumes
  and lattice spacings. The solid lines are the fits by the function~(\ref{fit_fun}).}
  \label{fig:condensates_ext_fields}
\end{figure}

Our numerical simulations reveal a gradual enhancement of the chiral condensate as the external magnetic field increases, Fig.~\ref{fig:condensates_ext_fields}. Qualitatively, this numerical result confirms various theoretical estimations~\cite{Agasian:Fraga:08:1:2:3,Gusynin:95:1,Smilga:97:1,Cohen:2007bt,Ebert:99:1,Zayakin:08:1}.

\begin{figure}
\begin{center}
  \includegraphics[width=6cm, angle=-90]{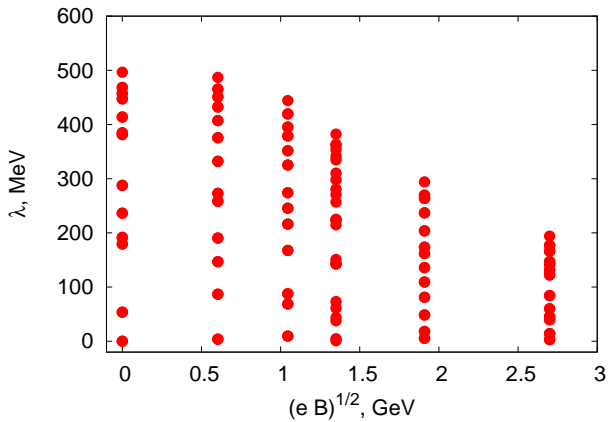}
\end{center}
  \caption{The $12$ lowest eigenvalues of the Dirac operator for typical configurations of non-Abelian gauge fields vs. the strength of external magnetic field for a selected configuration with unit topological charge. This configuration contains one zero Dirac eigenmode. }
  \label{fig:eigenvalue_flow:a}
\end{figure}

\begin{figure}
\begin{center}
  \includegraphics[width=6cm, angle=-90]{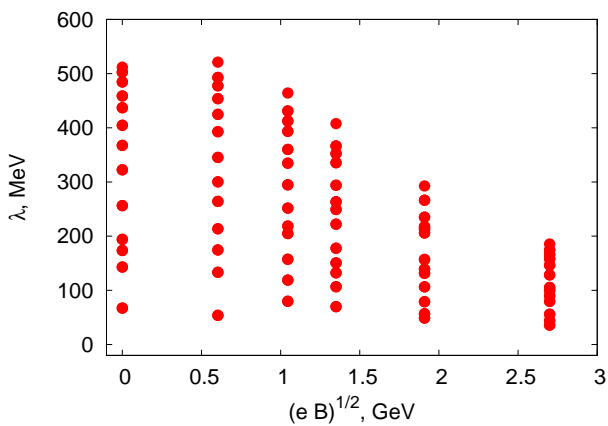}
\end{center}
  \caption{The same as in Figure~\ref{fig:eigenvalue_flow:a} but for a configuration with zero topological charge (no zero Dirac eigenmodes).}
  \label{fig:eigenvalue_flow:b}
\end{figure}

The enhancement of the condensate can be traced already in the behavior of Dirac eigenvalues. In Figs.~\ref{fig:eigenvalue_flow:a} and \ref{fig:eigenvalue_flow:b} we plot the dependence of twelve lowest Dirac eigenvalues for two typical configurations of non-Abelian gauge fields on the strength of the magnetic field. The stronger the field, the smaller eigenvalues of the Dirac operator and the larger the density of the near-zero modes. The later indicates the enhancement of the chiral condensate according to the Banks-Casher relation (\ref{BanksCasher}). Note also that the number of zero eigenvalues of the Dirac operator remain intact in external magnetic field, Fig.~\ref{fig:eigenvalue_flow:a}. Indeed, the Atyah-Singer theorem implies that the number of zero Dirac eigenmodes is equal to the topological charge of the gauge field configuration (we always have either left of right modes). The Abelian fields do not change the topology of the gauge fields, so that they do not influence the number of the zero modes.

We find that the dependence of the chiral condensate on the magnetic field strength can qualitatively be described by the analytical prediction (\ref{cc_vs_B_chPT}). To this end we fit our numerical data by the linear function
\begin{eqnarray}
\label{fit_fun}
\Sigma\lr{B} = \Sigma_0\,\lr{1 + \frac{e B}{\Lambda_B^2}}\,,
\end{eqnarray}
where $\Sigma_0 \equiv \Sigma\lr{0}$ and $\Lambda_B$ are fitting parameters. According to Figure~\ref{fig:condensates_ext_fields}
the function~(\ref{fit_fun}) reproduces our data for each set of lattice parameters quite well.
The the best fit results are given in the last two columns of Table~\ref{tbl:parameters}.

Before proceeding further we would like to mention a few important cautionary remarks.
Firstly, our weakest magnetic fields are still quite strong as one can see from Table~\ref{tbl:parameters}.
Indeed, the strength of the minimal magnetic field, $(e B_\mathrm{min})^{1/2}$, is still greater than
the scale imposed by the zero-field chiral condensate, $\Sigma^{1/3}_0$. In this regime the
prediction~(\ref{cc_vs_B_chPT}) of Ref.~\cite{Smilga:97:1} should not work, in general, as the
linear behavior is expected to be realized for much weaker fields. Thus, the linear behavior is
an unexpected result of our numerical simulations.

Secondly, the zero-temperature dependence of the chiral condensate $\Sigma\lr{B}$ on the magnetic fields slightly deviates
from a linear behavior at our weakest fields, as one can see in Figure~\ref{fig:condensates_ext_fields}. Despite deviations
are within error, the nonlinear features at weaker fields are not excluded by our data.

Thirdly, our largest fields are much stronger than the chiral scale, $(e B_\mathrm{max})^{1/2} \ll \Sigma^{1/3}_0$.
Yet, for the strongest magnetic fields that we use we do not see powerlike   $\Sigma\lr{B} \propto (e B)^{3/2}$
enhancement, which should be valid on dimensional grounds~\cite{Smilga:97:1}. The reason may lie in the fact that we are
studying the quenched theory, in which the virtual charged pions are absent, and the asymptotical behavior $e B)^{3/2}$
may be invalid until higher energies. However, the observed linear enhancement of the chiral condensate even in
the absence of the pion loops is intriguing feature of the non-Abelian gauge theory.

At zero temperature we used three different sets of lattice data to calculate the field dependence~(\ref{fit_fun}).
The data indicates that the finite lattice spacing and finite volume corrections to our results are quite small in the studied range of
parameters. The values for the zero field condensate, $\Sigma_0$, and for the slope parameter $\Lambda_B$, calculated at different
lattice volumes and spacings are quite close to each other. For further reference we take the values at largest lattice with finest lattice
spacing at $T=0$:
\begin{eqnarray}
\label{fit_best}
\Sigma^{\mathrm{fit}}_0 {=} {[\lr{320 \pm 5} \, \mbox{MeV}]}^3\,,
\qquad
\Lambda_B^{\mathrm{fit}} {=} (1.53 \pm 0.11) \, \mbox{GeV}\,.
\end{eqnarray}

The numerical value~(\ref{fit_best}) of the chiral condensate at zero magnetic field and at zero temperature
$\Sigma^{\mathrm{fit}}_0$ agree well with other numerical estimations in quenched $SU(2)$ gauge
theory~\cite{Hands:1990wc}. The numerical value of the slope
$\Lambda_B$ -- also given in Eq.~(\ref{fit_best}) -- is relatively close to the $T=0$ result of the
chiral perturbation theory~(\ref{cc_vs_B_chPT}), Ref.~\cite{Smilga:97:1}:
\begin{eqnarray}
\label{c:theory}
\Lambda_B^{\mathrm{th}} = \frac{4 \pi F_{\pi}}{\sqrt{\ln{2}}} = 1.97 \, \mbox{GeV}\,.
\end{eqnarray}
The later fact indicates that the quenched vacuum (studied numerically in this article) approximates the (electro)dynamics of
QCD with two massless quark flavours (used in theoretical analysis of Ref.~\cite{Smilga:97:1}) surprisingly well.
Thus, the quenched vacuum can qualitatively describe the enhancement of the chiral symmetry breaking induced by the
external magnetic field. Quantitatively, the quenching effects amount to be of the order to $30\%$.

Strictly speaking, the chiral condensate in quenched theory grows
logarithmically with lattice volume~\cite{Damgaard:2001xr}:
\begin{eqnarray}
\label{sigma_vs_vol}
\Sigma\lr{V} = \Sigma_{0}\left[1 + A \ln\frac{V}{V_{0}}\right]\,,
\end{eqnarray}
where $V_0$ is a certain characteristic volume.
Numerically, the logarithmic divergence in Eq.~(\ref{sigma_vs_vol}) is too small to be seen
in lattice data. Indeed, in our case changing the lattice volume by a factor of two
changes the chiral condensate only within the range of statistical
errors (see Table~\ref{tbl:parameters}). However, this fact raises a principal
question about establishing the correspondence between our values of $\Sigma$
and $\Lambda_{B}$ and the corresponding parameters in the full theory.
Numerically, we see that at our lattice volumes the chiral condensate
is quite close to its value, $\Sigma^{1/3} = 276(11)(16)\, \mbox{MeV}$
in the full theory~\cite{ref:Lang}. This means that the coefficient $A$
in (\ref{sigma_vs_vol}) is quite small~\cite{Damgaard:2001xr} and the volume
$V_{0}$ is close to our lattice volumes shown in Table~\ref{tbl:parameters}. The origin of
the logarithmic volume divergence in (\ref{sigma_vs_vol}) is the logarithmic divergence
in the one-loop correction to pion self-energy~\cite{Smilga:97:1,Damgaard:2001xr}.
From the Euler-Heisenberg lagrangian for a pion in external
electromagnetic field~\cite{Smilga:97:1} it is easy to see that the derivative
of $\Sigma$ over $B$ does not contain this divergence. Thus, one can
expect that the derivative
\begin{equation}
\label{eq:partial}
\frac{\partial \Sigma\lr{e B}}{\partial (e B)} = \frac{\Sigma_0}{\Lambda_B^2}\,,
\end{equation}
is finite in the infinite-volume limit. We see that actually our quenched result for
the slope coefficient $\Lambda_{B}$ is quite close to the analytical
prediction~(\ref{c:theory}) for the full theory~\cite{Smilga:97:1}.

The finite temperature reduces both the value of the chiral condensate $\Sigma_0$ and increases the value of $\Lambda_B$,
Table~\ref{tbl:parameters}. The last fact indicates that the linear dependence on the strength of the magnetic field gets
weaker with increase of temperature.

Concluding this Section we notice that in Ref.~\cite{ref:chromomagnetic}
the effects of the external diagonal gluonic fields on the phase diagram of QCD were studied.
These fields -- which are called the Abelian chromomagnetic fields -- turn out to enhance the chiral
condensate similarly to the external Abelian magnetic field of ordinary electromagnetism. Thus,
the strong magnetic fields of gluons and the strong magnetic fields of photons
have a similar influence on the condensate.

\subsection{Geometric structure of Dirac eigenmodes}

There are indications that chiral symmetry breaking is
closely related with the dominance of certain singular field
configurations in the vacuum of gauge theories, namely, with
center vortices and Abelian monopoles~\cite{Bornyakov:2007fz}.
(for a review of monopoles and vortices see, e.g., Ref.~\cite{reviews}).

It has been demonstrated that removal of
center vortices from gauge field configurations destroys both
confinement and chiral symmetry breaking \cite{deForcrand:99:1}.
More detailed studies revealed that near-zero eigenmodes of Dirac
operator, which, according to the Banks-Casher formula
(\ref{BanksCasher}), are responsible for the emergence of the chiral
condensate, are strongly correlated with center vortices
\cite{Polikarpov:05:2}. Similarly to center vortices, near-zero
eigenmodes of Dirac operator form complex fractal structures with
Hausdorff dimensionality between 2 and 3 \cite{Polikarpov:05:2,Aubin:04:1}.
Thus, it is interesting to study the effect of background magnetic fields
on the localization of low-lying Dirac eigenmodes.

\begin{figure}[ht]
\begin{center}
  \includegraphics[width=6cm, angle=-90]{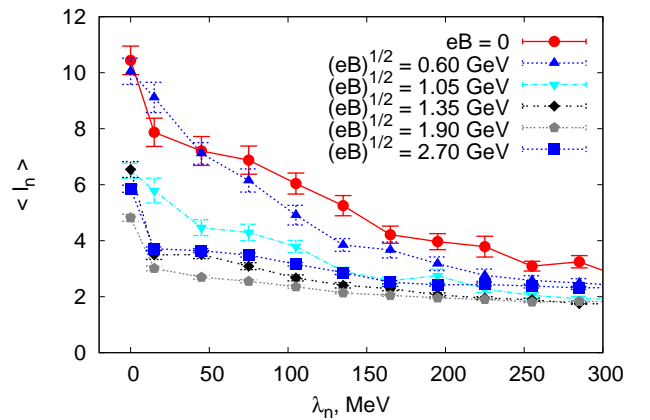}
\end{center}
  \caption{IPR~(\ref{IPR}) of Dirac eigenmodes with $\lambda_n< 500 \, \mbox{MeV}$ for different strengths of the background magnetic field.}
  \label{fig:ipr}
\end{figure}

 A commonly used measure of localization is the Inverse Participation Ratio (IPR):
\begin{eqnarray}
\label{IPR}
I_{n} = V \int {\mathrm d}^{4} x \, \rho_{n}^{2}\lr{x}
\quad
\mbox{with}
\quad
\int {\mathrm d}^{4}x \, \rho_{n}\lr{x} = 1\,.
\end{eqnarray}
Here $\rho_{n}\lr{x} = \psi^{\dag}_{n}\lr{x}\, \psi_{n}\lr{x}$ is the density of the $n$-th Dirac eigenmode and $V$ is the volume of the system.

The larger is the IPR, the more localized is the eigenmode $\psi_{n}$. For the function localized in a single point, $\psi\lr{x} \sim \delta^{(4)}\lr{x-x_{0}}$, the IPR is equal to the total volume of space $V$, while for a delocalized function, $\psi\lr{x} \sim 1/\sqrt{V}$, the IPR is equal to unity. The dependence of the IPR on the magnetic field strength and the eigenvalue $\lambda_{n}$ is illustrated in  Fig. \ref{fig:ipr}. For not very strong field strengths, magnetic field decreases IPR and thus delocalizes Dirac eigenmodes.
\begin{figure}[ht]
  \includegraphics[width=95mm,clip=true]{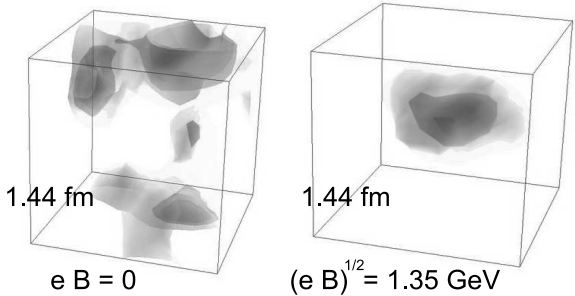}
  \caption{Three-dimensional cuts of the density of a low-lying Dirac eigenmode
  for the same configuration of non-Abelian gauge fields are plotted at zero magnetic field (left) and
  at $\sqrt{e B} = 1.35 \, \mbox{GeV}$ (right).}
  \label{fig:eigenmodes}
\end{figure}

In order to visualize how the structure of the eigenmodes of the
Dirac operator is modified in the external magnetic field,
in Fig.~\ref{fig:eigenmodes} we have shown the three-dimensional cuts of
the four-dimensional level surfaces of the densities of two
near-zero Dirac eigenmodes for the same typical configuration of
non-Abelian gauge fields with background magnetic field equal to
zero (on the left) and with $\sqrt{e B} = 1.35 \, \mbox{GeV}$ (on
the right). While at zero fields the eigenmode indeed has a highly
irregular structure and stretches through the whole space, at
$\sqrt{e B} = 1.35 \, \mbox{GeV}$ it occupies a compact region and
has more or less smooth level surfaces. However, at zero field the
IPR is still significantly higher. Why, on the one
hand, the magnetic field seemingly increases the localization, and,
on the other hand, decreases the IPR? This apparent paradox can be
resolved if one remembers that at zero magnetic field near-zero
eigenmodes of the Dirac operator are localized on lower-dimensional
irregular manifolds \cite{Aubin:04:1}, which are strongly correlated
with the worldsheets of center vortices \cite{Polikarpov:05:2}.
These manifolds percolate and thus fill all the physical space, in
the sense that any two arbitrarily distant points have a non-zero
probability to belong to the same vortex worldsheet. The total
volume of these manifolds is, however, zero, and their IPR diverges
as some power of ultraviolet cutoff \cite{Polikarpov:05:2,Aubin:04:1}. If, on the other hand, the eigenmodes are
localized in finite four-volumes, their IPR remains finite in the
continuum limit.

In a non-Abelian gauge theory only quarks interact
with background Abelian fields, and in quenched theory the
path-integral weight for the gauge fields $A_{\mu}$ is not modified.
In particular, background fields do not affect positions of center
vortices. Since the configurations of center vortices are the same
for both plots in Fig. \ref{fig:eigenmodes}, we have to conclude
that in background magnetic field near-zero Dirac eigenmodes become
less correlated with center vortices. For sufficiently strong fields
these modes are no longer associated with lower-dimensional
structures with typical size set by the ultraviolet cutoff scale,
but are rather localized on a physical scale determined by both
hadronic scale and the strength of magnetic field.

The fact that at some large $\sqrt{e B}$ ($\approx 2 \, \mbox{GeV}$ in our case)
the IPR starts growing with magnetic field can be explained as follows:
at this magnetic field and at finite lattice spacing the total
volume of lower-dimensional manifolds on which the eigenmodes can be
localized exceeds the localization volume set by the magnetic field.
Then this effect should disappear in the continuum limit, $a \rightarrow 0$.
One the other hand the increase of the IPR at high magnetic fields may have
another simple explanation. At very strong fields the structure of the quark eigenmodes
should dominantly be determined by the magnetic field and not by the underlying non-Abelian gauge field.
As the magnetic field increases the quark eigenmodes tend to get localized in the transverse
(to the field) directions filling out low Landau levels. The radius of the Landau levels shrinks as
the magnetic field gets stronger and this effect may lead to the observed increase in the IPR. However,
this effects operates at very strong gauge fields, at which the chiral condensate should presumably
be determined by the scale of the magnetic field alone~\cite{Smilga:97:1}, $\Sigma \sim (e H)^{3/2}$.

The effect of delocalization of near-zero eigenmodes in the background magnetic field can also
qualitatively  explain why the chiral condensate increases with the field strength. Weaker
correlation between center vortices and near-zero Dirac eigenmodes implies an increase of
the total volume of space where these modes can be localized. Literally, high magnetic fields
tear away the eigenmodes from the center vortices. Thus, near-zero modes become
more degenerate~\footnote{In a free theory with background magnetic field near-zero modes
can be localized in any point of physical space, and the chiral condensate (\ref{BanksCasher})
indeed diverges in the continuum limit \cite{Gusynin:95:1, Wiese:08:1}.} and the density of
near-zero eigenvalues $\rho\lr{\lambda \to 0}$ increases, implying growth of the chiral
condensate~(\ref{BanksCasher}).

\section{Conclusions}

Summarizing, we conclude that our numerical simulations
show a clear {\it qualitative} evidence that background magnetic field enhances spontaneous breaking
of chiral symmetry in a non-Abelian gauge theory. We confirm that the chiral condensate rises
linearly with increase of the field strength~(\ref{fit_fun}), in at least qualitative agreement with
the prediction~(\ref{cc_vs_B_chPT}) of the chiral perturbation theory~\cite{Smilga:97:1}. The linear
dependence of the condensate on the magnetic field is governed by the parameter~$\Lambda_B$. Surprisingly
we have found that at zero temperature the value of the parameter~$\Lambda_B$ of the quenched
theory~(\ref{fit_best}) is close to the analytical prediction of the chiral theory~(\ref{c:theory}).

We argue that the slope of the linear dependence of the chiral condensate on the strength of the magnetic
field~(\ref{eq:partial}) is not affected by the logarithmic volume dependence specific to the quenched
limit~\cite{Damgaard:2001xr}.

The linear dependence of the chiral condensate on the strength of the magnetic field is an unexpected
result of our study because our fields are much stronger than the fields used in the analytical investigation of
Ref.~\cite{Smilga:97:1}. Moreover, the enhancement of the chiral condensate in QCD is caused by the
coupling of the external magnetic field to the charged pion loops~\cite{Smilga:97:1}, which are absent
in the quenched approach used in this paper.
Our result indicates that non-Abelian gauge fields alone play a significant role in the enhancement of
the chiral condensate at nonzero magnetic fields. The dominance of this contribution may be revealed
in further lattice simulations with dynamical fermions.

We found that the dependence of the chiral condensate on the magnetic field -- encoded in
the parameter $\Lambda_B$ -- becomes much less pronounced with increase of the temperature.

We observe that the (de)localization properties of near-zero Dirac eigenmodes are crucial for the
enhancement of chiral symmetry breaking in the external magnetic field. Our results suggest that the
microscopic mechanism of the enhancement is the delocalization of the near-zero eigenmodes off the
center vortices which increases the degeneracy of near-zero modes.

\section*{Acknowledgments}

The authors are grateful to V.~G.~Bornyakov and S.~M.~Morozov for interesting discussions. This work was partly supported by Grants RFBR Nos. 06-02-04010-NNIO-a, 08-02-00661-a, 06-02-17012, 09-02-00629-a, and DFG-RFBR 436 RUS, BRFBR F08D-005, a grant for scientific schools Nos. NSh-679.2008.2 and NSh-4961.2008.2, by the Federal Program of the Russian Ministry of Industry, Science and Technology No. 40.052.1.1.1112, by the Russian Federal Agency for Nuclear Power, and by the STINT Institutional grant IG2004-2 025.  P. V. Buividovich is also partially supported by the Euler scholarship from DAAD, by the scholarship of the ``Dynasty'' foundation and by the grant BRFBR F08D-005 of the Belarusian Foundation for Fundamental Research.
The calculations were partially done on the MVS 50K at Moscow Joint Supercomputer Center.

\end{document}